\documentstyle[aas2pp4]{article} %preprint style
\newcommand{\bee}{\begin{eqnarray}}
\newcommand{\ene}{\end{eqnarray}}
\input epsf.sty
\begin{document}
\title{Does the Sun Shrink with Increasing Magnetic Activity?}
\author{ W. A. Dziembowski
$^{1,2}$, P. R. Goode$^{2}$, and J. Schou$^3$ }
\affil{ $^1$ Warsaw University Observatory and Copernicus Astronomical Center,
Poland\\
$^2$ Big Bear Solar Observatory, New Jersey Institute of
Technology, U.S.A\\
$^3$ W.W.Hansen Experimental Physics Laboratory, Stanford University,
U.S.A.}
\begin{abstract}

We have analyzed the full set of SOHO/MDI f- and p-mode oscillation
frequencies from 1996 to date in a search for evidence of
solar radius evolution during the rising phase of the current activity
cycle.  Like Antia et al. (2000), we find that a significant fraction
of the f-mode frequency changes scale with frequency; and that if
these are interpreted in terms of a radius change, it implies a shrinking
sun.  Our inferred rate of shrinkage is about 1.5 km/y, which is somewhat
smaller than found by Antia et al.  We argue that this rate does not refer
to the surface, but rather to a layer extending roughly from 4 to 8 Mm
beneath the visible surface.  The rate of shrinking may be accounted for
by an increasing radial component of the rms random magnetic field at
a rate that depends on its radial distribution.  If it were uniform, the
required field would be $\sim7$ kG.  However, if it were inwardly increasing,
then a 1 kG field at 8 Mm would suffice.

To assess contribution to the solar radius change
arising above 4Mm, we analyzed the p-mode data.
The evolution of the p-mode frequencies may be explained by a magnetic
field growing with activity.  Our finding here is very similar to that
of Goldreich et al. (1991).  If the change were isotropic, then 
a 0.2 kG increase, from activity minimum to maximum, is required at the
photosphere, which would grow to about
1 kG at 1 Mm. If only the radial component of the
field were to increase, then the requirement for the photospheric
field increase is reduced to a modest 60-90 G.
A relative decrease in temperature of the order of $10^{-3}$
in the sub-photospheric
layers, or an equivalent decrease in the turbulent energy, would have
a similar effect to the required inward growth of
magnetic field change.

The implications
of the near-surface magnetic field changes depend on the anisotropy of
the random magnetic field.  If the field change is predominantly radial,
then we infer an additional shrinking at a rate between 1.1-1.3 km/y at the
photosphere.  If on the other hand the increase is isotropic, we find
a competing expansion at a rate of 2.3 km/y. In any case,
variations in the sun's radius in the activity cycle
are at the level of $10^{-5}$
or less, hence have a negligible contribution to the
irradiance variations.

\end{abstract}

\section{Introduction}

Measuring the sun's radius, and its variability, are significant,
long-standing problems, especially in the context of understanding
the cause of solar irradiance variations.  Recently, it has been pointed out
that helioseismology can provide a useful measure of the solar radius.
Schou et al. (1997) and Antia (1998) showed that f-modes
frequencies are good probes of the radius, and they inferred a value
of the solar radius which is about 300 km smaller than the one adopted
in solar models at that time.  The model values were based on a
direct measurement of the sun's photospheric radius.
The smaller radius has been confirmed by Brown and
Christensen-Dalsgaard (1998) from many years of transit measurements using
the Solar Diameter Monitor.  The connection between the ``true'' solar radius
and that inferred from f-modes is explained in Section 3.1.

Following the suggestion of Schou et al. (1997),
Dziembowski et al. (1998, 2000)
used f-mode data from the MDI/SOHO instrument to determine the evolution
of the seismic solar radius through the rising phase of the present activity
cycle. They reported statistically significant variations that showed no
apparent correlation with activity measures.  On the other hand,
with GONG f-mode frequencies (ranging between 1.015 mHz to 1.425 mHz, or
equivalently $\ell$ from 100 to 200), covering the rising phase of
activity to the beginning of 1999, Antia et al. (2000)
found a net decrease of about 5 km in the solar radius.
They attributed the difference with Dziembowski et al. (1998) to the
latter's use of higher degree modes (up to $\ell$=300).  They pointed
out the latter $\ell$-modes are more likely to be effected by factors other
than an evolving radius.

In the present work, we use oscillation data from SOHO/MDI
covering 1996.3-2000.5.  We first use f-mode data in an effort to
infer a signal of radius change.   We then modify our earlier
analyzes to consider other sources of the variations. Our analysis
is preceded by an explanation of the meaning of the ``seismic''
radius inferred from f-modes.  We then compare our results to
those of Antia et al., and interpret the inferred rates in terms
of magnetic field and temperature changes.

Our interpretation of p-mode frequency changes is predicated on the
work of Goldreich et al. (1991) who analyzed BBSO measurements
from the rise of the previous cycle (cycle 22).  We use our inference
on the behavior of the sub-photospheric layers to constrain radius
changes arising there.

\section{Frequency data from SOHO/MDI}

In the present study, we use 19 MDI data sets containing centroid frequencies
determined from measurements made between May 1, 1996 and June 21, 2000,
with a break between June 16 and October 22, 1998, when there was no contact
with SOHO. The sets are typically 72-days long, except those immediately
before and after the break, which are shorter.
The centroid frequencies, $\nu_{\ell,n}$, were determined by
the method described by Schou (1999).

The sets contain between 112 and 203 f-mode frequencies,
with earlier sets having more data. The maximum $\ell$-value is 300 and
the minima range from 89 to 137.  The number of p-mode frequencies range
from 1589 to 1906.  Again, the earlier sets are mode abundant. The
p-modes range between $\ell$=0 and 200. The differences in mode
composition are not important in the case of p-modes.  The number of
overlapping modes is large enough for a detailed study of frequency
changes. In the case of
f-modes, the difference in the $\ell$-range may be important, and therefore in
our study of the solar radius changes, we used only modes with $\ell\ge137$.

\section{Inferences from f-mode frequency changes}

\subsection{Helioseismic radius}

All helioseismic determinations of the solar radius to date have relied on
the following asymptotic relation for f-modes frequencies,
\begin{equation}
{\Delta\nu_\ell\over\nu_\ell}=-{3\over2}{\Delta R\over R}.
\end{equation}
Antia et al. (2000) pointed out that using this relation
for modes with $\ell$ extending up to 300, as Dziembowski et al. (1998) did,
is not justified because of significant departures from
$\nu\propto R^{-1.5}$ are present in higher $\ell$'s.
The departure increases with $\ell$, which as
Brown (1984) first suggested could be accounted for as an effect of
turbulence in the upper convective zone.
Detailed models of this effect have been developed by Murawski \& Roberts
(1993a, 1993b). (For the most recent work on the subject see M{\c e}drek \&
Murawski 2000). However, surface magnetic fields may also have significant
effects on f-mode frequencies (Evans \& Roberts, 1990; Jain \& Roberts).
With these two sources of perturbation to f-mode frequencies,
we must contemplate solar cycle changes beyond that of a simple radius change.
The relative contribution of the near-surface changes are expected to increase
with $\ell$, because such changes should be inversely proportional to
mode inertia, $I_\ell$, which sharply
decreases with $\ell$.

There is another problem in applying Eq.(1) in a search for
the radius variations correlated with activity.  This problem follows from
the fact that the induced modifications are quite non-uniform, and each
f-mode has it is own radius, $R_\ell$, which is given by
\begin{equation}
R_\ell=\left({1\over I_\ell}\int r^{-3} d I_\ell\right)^{-1/3}.
\end{equation}
With this definition, we get from the variational principle for
oscillation frequencies (see Appendix)
\begin{equation}
\nu_l={1\over2\pi}\sqrt{(L-2){GM\over R_\ell^3}},
\end{equation}
where $L=\sqrt{\ell(\ell+1)}$.
This is a very accurate expression.
The relative departures from
equality range from $2\times10^{-4}$ at $\ell=100$ to
$8\times10^{-5}$ at $\ell=300$. In a linear approximation in
terms of $(R-R_\ell)$, this is the same as the formula obtained by
Gough (1993).

For high degree modes, the f-mode radii are close to the solar radius.
The values of $R_\ell/R$ range from 0.9883 at $\ell=100$
to 0.9946 at $\ell=300$. While we have $R_\ell\approx R$, a
corresponding approximation for $\Delta R_\ell$ is quite problematic. When the
f-mode frequencies were used to refine the value of the radius for modeling
the sun, we could expect an approximate, homologous relation,
$R_\ell\propto R$.  But such a relation cannot be expected in the case
of the activity induced changes, which we believe to be confined to
the outermost part of the sun. If the
data show that $\Delta\nu_\ell\propto\nu_\ell$, as Antia et al.(2000) found,
then the simplest interpretation is that, indeed we have
$\Delta R_\ell$ being constant over the range of considered $\ell$-values.
Then, the inferred value of $\Delta R$ in Eq.(1) refers to
the range of depths beneath the photosphere corresponding to
the range of $\ell$'s in the data sets. Antia et al. (2000) used
modes in the 100--200 range, which translates to 10--6 Mm in depth.
Their finding implies that this layer was moved downward
by about 5 km during the two years they considered.  The truth is, with
these data, we
cannot say anything about what happened in the layers above.  Thus, we have no
information about the evolution of the photospheric radius of
the sun.

\subsection{Formal determination of the rate of shrinking from f-modes}

To account for the effect of the near-surface changes on f-mode frequencies
and possible differential changes,
we modified Eq.(1) into
\begin{equation}
\Delta\nu_\ell=-{3\over2}{\Delta R_f\over R}\nu_\ell+
{\Delta\gamma_f\over I_\ell},
\end{equation}
where $\Delta R_f$ denotes the radius change inferred from
a particular set of f-modes.
For the calculation of $I_\ell$, we adopted the following normalization
of the eigenfunctions
\begin{equation}
(\xi_r)_{\rm ph}={2\times10^4\over\sqrt{\bar\rho R^5}}Y^m_\ell,
\end{equation}
where $\xi_r$ denotes radial displacement of the fluid element.
With such a normalization, the values of $I_\ell(\nu)$ are of the order
of unity for p-modes in the 2--4 mHz range. For the f-modes, the $I_\ell$
values are significantly larger (see Table 1).

We determined $\Delta R_f$ and $\gamma_f$ by a least-squares fitting.
In Fig. 1, we show values of  $\Delta\nu_\ell$ for selected data sets.
The first two sets were obtained at solar minimum. The 1999.4 set was
taken near the middle of the phase of rising activity, and the last is at the
current maximum. Here, $\Delta$ denotes the difference between the
solar data and the reference model.
The reference solar model is that of Christensen-Dalsgaard et al.(1996).
It has the same radius used by Dziembowski et al. (1998). That is, the model
uses the ``old'', too large value of the solar radius (not that of Brown and
Christensen-Dalsgaard 1998), and this is why the frequency differences
are rather large.  A small difference in
the reference model radius is inconsequential for the inferred temporal
changes.
\begin{figure}
\epsfxsize=1\hsize
\epsffile{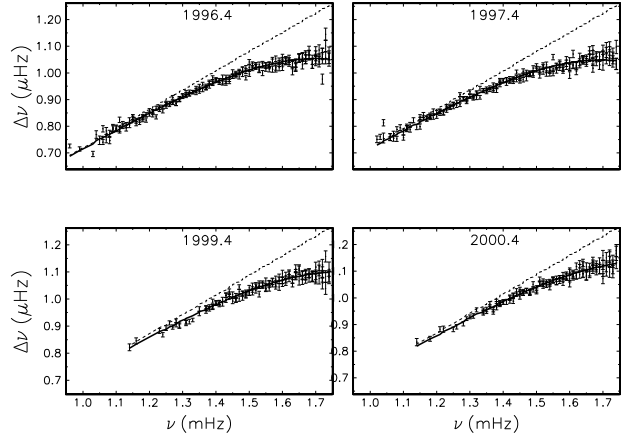}
\caption{Differences between measured and calculated f-mode frequencies.
The error bars show estimated standard deviations of measured
values.  The dates correspond to the center of the individual
72-day long measurement
periods. The solar model was calculated assuming $R_\odot=695.991$ Mm.
The solid line represents the fit to Eq. (4). The dashed straight line
represents
the part attributed to the difference between the solar radius and that
adopted in the model.}
\end{figure}

We assumed that $\Delta\gamma_f$ is $\nu$-independent, and as we see in
Fig. 1, the solid line is a good fit to the data.
The $\chi^2$'s vary from 1.2 (2000.4 set) to 1.84 (1996.4), except
for the significantly
worse fit ($\chi^2=3.55$) found for the 1998.9 set, which was
the first one taken after the recovery of
SOHO. We also tried fitting $\gamma_f$ as a low-order
polynomial depending on frequency, but this did not improve the fit.

We see the departure from the linear relationship implied by the
the radius adjustment sharply increases with $\nu$.
Antia et al. (2000) considered only modes with $\nu<1.44$ mHz, and
it seems that the departure from a straight line is still small.
However, this is somewhat misleading because we used a model with
much too large a radius.  As we shall see in Table 1, at the level of
changes of a few nanohertz (i.e. radius changes of a few km), the
difference is quite significant.  We emphasize that high $\ell$-modes
are important because with increasing $\ell$, $R_{\ell}$
approaches the solar radius.  For such modes, including
$\gamma_f$ is essential, which implies that we have to rely on
Eq.(4) rather than Eq.(1). With Eq.(1), we get a much poorer
fit ($\chi^2=4.4 - 16.5$) and the correction to the solar radius is larger
by some 20 km.  This illustrates the trade-off -- increasing $\ell$
moves us closer to the surface, but such high $\ell$'s are more strongly
contaminated.

\begin{figure}
\epsfxsize=0.7\hsize
\epsffile{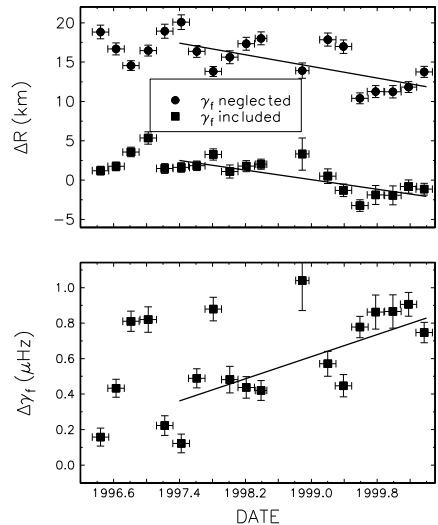}
\caption{{\it Upper panel}: Variation of solar radius between 1996.4 and
2000.4 inferred from f-mode frequencies with and without the $\gamma_f$-term.
Two straight lines represent linear fits to the data starting from 1997.4
when the rise of cycle 23 began. {\it Lower panel}: Corresponding
variation of $\gamma_f$, which describes remaining near-surface contribution
to f-mode frequency variations.}
\end{figure}

\begin{table}
\caption{Contributions to f-mode frequency shifts during the rising phase of
cycle 23 }

\medskip

\begin{tabular}{ccccc}
\hline $\ell$ &  $\nu_\ell [{\rm mHz}]$ & $I_\ell$ & $\Delta\nu_R [\mu{\rm Hz}]$
& $\Delta\nu_\gamma [\mu{\rm Hz}]$\\
\hline
 100 & 1.02 & 381 & 0.010 & 0.0012\\
 130 & 1.15 & 165 & 0.011 & 0.003\\
 200 & 1.43 & 39. & 0.014 & 0.012\\
 300 & 1.74 & 9.4 & 0.017 & 0.050\\
\hline
\end{tabular}
\end{table}

%to be removed it:
%Our data sets from solar minimum contain some modes with $\ell$ as
%low as 90.  To avoid spurious changes in $\Delta R_f$, following from
%the difference in the minimum values of $R_\ell$, we eliminated all the data
%for which $\ell\le$137.]]

In Fig. 2, we show the variations of the solar radius and $\gamma_f$
inferred from f-modes from the truncated data sets.
The rise of the current activity
cycle began in 1997.4 which was marked by a sharp rise of the seismic activity
indicators (Dziembowski et al. 1998). A corresponding sharp rise of p-mode
frequencies beginning at this time may be seen in Fig. 3 here.
That is why we choose 1997.4 to begin our linear fits.
We have no explanation, as yet, for the relatively large fluctuations
in $\Delta R_f$ which appear to have a one-year period.
For comparison, we
also show the result obtained when the $\gamma_f$-term is ignored. There is
a difference, but not as large as one might anticipate by looking at Fig.1.
The rate of radius decrease is only insignificantly higher than in our standard
version, and the error is larger.

In detail, we found from our linear fit, with the $\gamma_f$,
\begin{equation}
{d R\over dt}=(-1.51\pm 0.31)\quad{\rm km/y},
\end{equation}
and without the $\gamma_f$-term,

$${d R\over dt}=(-1.82\pm 0.64)\quad{\rm km/y}.$$
The values are similar to those found by
Antia et al. (2000). To make a closer comparison, we
truncated our data sets at $\ell=200$, and then we found
$${d R\over dt}=(-1.80\pm 0.38)\quad{\rm km/y}.$$
Having in mind that we still miss modes between $\ell=100$ and 137,
it is fair to say that there is no disagreement between our findings
and theirs, implying that at a depth of from 6 to 10 Mm the sun
shrank  by some 4 to 6 km during the rising phase of this activity cycle.

How reliable is this finding? The main concern is the role of the
near-surface perturbation and the cross-talk between the two
terms on the right hand side of  Eq.(4).
In the lower panel of Fig. 2, we show the $\gamma$'s.
The linear fit for $\gamma$, which is visibly poorer, yields
\begin{equation}
{d\gamma_f\over dt}=(0.180\pm 0.051)\quad\mu{\rm Hz/y}.
\end{equation}
The relative contribution of the two terms to overall f-mode frequency
variations
depends on $\ell$. In Table 1, we compare these two contributions,
denoted by$\Delta\nu_R$ and $\Delta\nu_\gamma$
for selected $\ell$-values. The increasing role
of $\Delta\nu_\gamma$ is a consequence of decreasing mode inertia.
It should be noted  that $\Delta\nu_\gamma$ yields an appreciable
contribution to $\Delta\nu$ even for modes with $\ell\le200$.
Caution is necessary, but we will proceed further assuming that
the effect is indeed real.

\subsection{Accounting for the rate of shrinkage}

Even as small as it seems, a shrinking of the sun's radius during
the rising phase of activity is not easy to explain.  To investigate,
we write the Lagrangian change of the local radius in the form
\begin{equation}
\Delta r(r_0) =r-r_0=-\int_{r_b}^{r_0}
{\Delta\rho\over\rho}\left({x\over r_0}\right)^2dx,
\end{equation}
where $r_b$ is the radius at the bottom of the layer perturbed
by activity, and $r_0$ is the radius at a specified fractional
mass, $M_r/M$, at activity minimum
and $\Delta\rho$ denotes the horizontally averaged change of density.
We obtain a more revealing form of Eq.(8) by expressing
$\Delta\rho$ in terms of the averaged entropy and magnetic field changes.

For the horizontally averaged gas pressure in the presence of a random
magnetic field we have, after Goldreich et al. (1991),
 \begin{equation}
\Delta P_g=-\Delta(\beta P_m),
\end{equation}
where
$$P_m={\overline{B^2_h}+\overline{B^2_r}\over8\pi}$$
is magnetic pressure and
$$\beta ={\overline{B^2_h}-\overline{B^2_r}\over8\pi P_m}$$
is a measure of the statistical anisotropy of the field.

With the use of thermodynamical relations, we determine
\begin{equation}
\Delta r=\int_{r_b}^{r_0}\left[{1\over\Gamma_1}{\Delta(\beta P_m)\over P_g}+
(-\rho_T){\Delta S\over c_p}\right]\left({x\over r_0}\right)^2dx,
\end{equation}
where $\rho_T$ denotes the logarithmic derivative of density at constant
pressure. The remaining thermodynamical quantities have their standard meanings.
At the relevant depths, the gas is nearly ideal. Thus, we may use
$\rho_T=-1$, ${1/\Gamma_1}=0.6$, and find
$${\Delta S\over c_p}={\Delta T\over T}-0.4{\Delta P_g\over P_g}.$$
The irradiance from an active sun is higher than average. If the same is true
about luminosity then we should have $\Delta S<0$. Hence, a negative
contribution to $\Delta r$. However, this must be very small. If  $\Delta S$
refers to the whole convective zone then a $10^{-3}$ luminosity increase
translates to an annual decrease in $\Delta S/c_P$ of $10^{-7}$.
Another possibility is an increase in the superadiabatic gradient,
$\nabla_{\rm con}-\nabla_{\rm ad}$, but this seems unlikely too.
The annual decrease of $\Delta R_f=1.5$ km refers
to the layer of $r/R=0.988 - 0.995$. Thus, $\Delta R_f$
must arise mostly beneath $r=R_{137}=0.988R$.
At this depth, according to a mixing-length model,
$\nabla_{\rm con}-\nabla_{\rm ad}\approx2\times10^{-4}$, which rapidly
decreases going inward. We would need an order of magnitude increase in the
superadiabatic gradient to account for our rate of shrinking.

A more acceptable explanation would be a variation in the magnetic field.
The consequences of  a
magnetic field increase depend on $\beta$. For a purely radial
field ($\beta=-1$), the increase implies contraction.
For an isotropic field ($\beta=1/3$) the increase implies expansion.
The field geometry implying the minimum increase to account for
the rate of the shrinking corresponds to $\beta=-1$. Then, we have
$\Delta\!<B\!>_{\rm rms}=\sqrt{\overline{\Delta(B_r^2)}}$,
and assuming a constant rate across the lower convective zone,
we infer
$${d\!<B\!>_{\rm rms}\over dt}\approx 7.2 {\rm kG/y}.$$
The value at $R_{137}$ may be reduced, for instance, to 1 kG/y if one
allows an exponential increase of the rate to about 43 kG/y at
the base of the convection zone.

Thus, what we have inferred from the f-mode frequency change is a
non-trivial constraint on the internal magnetic field change. Let us note that
if the field increase were predominantly isotropic, we would see an expansion
rather than a contraction. The field increase inferred from the residual
(after removing the near surface contribution) part of the p-mode frequency
change was  about 60 kG at 25-100 Mm (Dziembowski et al., 2000).
This high value could be consistent with the shrinking rate only if $\beta$ is
close to zero, that is if the field is essentially force-free, which is not
a likely possibility. Thus, we are now  skeptical about the
reality of that large field change we reported earlier.

Our inference regarding the solar radius change is limited by the lack of
accurate information about what happened in the outer 4 Mm of the solar
interior. This
is the region where we may expect the largest activity induced variations for
two reasons. First, the rapid decline of gas pressure and second,
the thermal structure of this layer is more susceptible to changes in
the efficiency of the convective energy transport induced by the field
changes.  The f-mode data we have at hand provide some information about
changes in this layer through  the $\gamma_f$.  Similar, but much more
accurate information is available in the p-mode data, which we now consider.

\begin{figure}
\epsfxsize=0.7\hsize
\epsffile{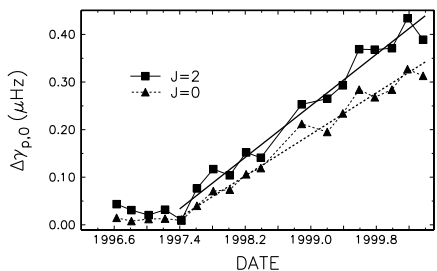}
\caption{Variation of the mean value of $\gamma$ with two versions
of its polynomial dependence inferred from p-mode frequencies.
As in the case of f-modes shown in Fig. 2, the
linear fit corresponds to the data starting from 1997.4
when the rise of cycle 23 began. The error bars would be within the
symbols.}
\end{figure}

\section{Inference from p-mode frequency changes}

\subsection{The near-surface source of the p-mode frequency changes}

The p-mode spectrum of MDI frequency data is about 13 times richer
than that for f-modes. Unfortunately, p-modes are not directly
useful for
determining changes in the solar radius. The simple relation,
$\nu\propto R^{-1.5}$ would be valid for p-modes only if the changes were
homologous throughout the whole sun.  This is far from the truth for the
changes we are considering here.  However, from p-modes one may make a much
more precise determination, than from f-modes,
of the near-surface perturbation.  For
p-modes, we call it $\Delta\gamma_p$, and it
describes frequency changes caused by a
variable perturbation localized near the surface.  In the present application,
however, taking into account the $\nu$ dependence is required
for an accurate fit. We express the dependence in the form of a
Legendre polynomial series with argument
$$s={\nu-(\nu_l+\nu_h)/2\over\nu_h-\nu_l},$$
where $\nu_l$ and $\nu_h$ denote the lowest and the highest
frequencies in the data set. Thus, we write
\begin{equation}
\Delta\nu_{\ell,n}={1\over I_{\ell,n}}\sum_0^J\Delta\gamma_{p,j}P_j(s).
\end{equation}
Here $\Delta$ is with respect to the 1996.4 data set.
The number $J$ was increased until $\gamma_{p,0}$ stabilized within the
errors and $\chi^2$ stabilized. This occurred for $J=2$.
In Fig. 3, we plot $\Delta\gamma_{p,0}$ for $J=0$ and 2.
Variations of $\gamma_p$ are indeed much more accurately determined than those
of $\gamma_f$.  For $J=2$, we find the rate
\begin{equation}
{d\gamma_{p,0}\over dt}=(0.149\pm 0.008)\quad\mu{\rm Hz/y}.
\end{equation}

The dependence of $\gamma(\nu)$ yields an important constraint on
the localization of the source of solar cycle variations in p-mode frequencies.

Following Goldreich et al. (1991), we link the frequency change
to the change of the mean squared magnetic field and a Lagrangian
change of a single thermodynamic parameter. For the latter, we prefer
to use temperature rather than entropy which was used by Goldreich et al.
From Eqs. (14) and (15) of Goldreich et al., we get
the following expression for the change of $\gamma_p$,

\bee
\Delta\gamma_p&=&{1\over8\pi^2\nu}\int d^3\vec x\vert{\rm div}\vec\xi\vert^2
\{P\Gamma_1(1+\Gamma_\rho)\rho_T{\Delta T\over T}\nonumber\\&&
+[1+\Gamma_1(\Gamma_P+\Gamma_\rho\rho_P)\nonumber\\&&
-\beta(\Gamma_1-1+\Gamma_1\rho_P)]\Delta P_m\}.
\ene
Here, we denote by $\Gamma_P$ and
$\Gamma_\rho$, the logarithmic derivatives of $\Gamma_1$.
The ideal gas equation cannot be used in the layers where most of the
contribution to $\Delta\gamma_p$ arises.

Goldreich et al. (1991) explained the
p-mode frequency changes during the rising phase of cycle 22 in
terms of magnetic field and temperature changes, with former being
dominant and causing the frequency increase. They invoked a chromospheric
temperature increase to explain the reversal in the increasing trend in
$\Delta(\nu)$. We do not see such a trend in our data. Thus, as a
first guess we interpret $\Delta\gamma_p$ in terms of magnetic field
changes. Later, we will discuss other sources of the p-mode
frequency changes.

We considered two values of $\beta$, -1 and 1/3,
and the following form for the depth, $D$, dependence
of magnetic field increase
$$\Delta<\!B\!>_{\rm rms}=\left\{
\begin{array}{ll}
B_b & \mbox{ if $D\ge D_b$}\\
B_b+\lambda\left({D-D_b\over D_b-D_m}\right) & \mbox{if $D_t<D<D_b$}\\
\Delta<\!B\!>_{\rm rms}(D_t) & \mbox{if $D\le D_t $}
\end{array}
\right.,$$
where $D_m=-0.485$Mm denotes $D$ at the temperature minimum,
and $B_b$, $D_b$, and $\lambda$ were determined by fitting the three
terms in the series given by Eq. (11).
For $D_t$ we adopted either $D_m$ or 0.

\begin{figure}
\epsfxsize=0.6\hsize
\epsffile{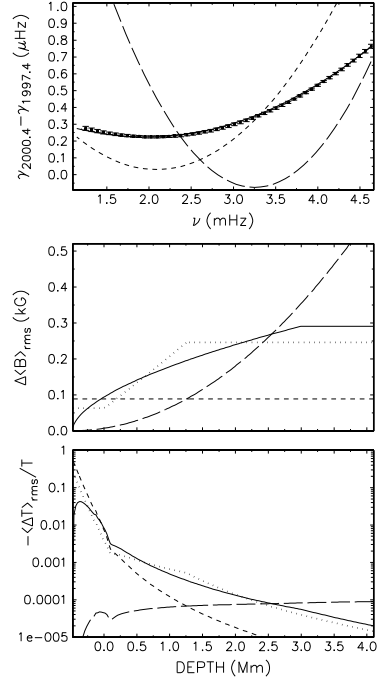}
\caption{ In the top panel, points with the error bars represent
$\Delta\gamma_0(\nu)$ inferred from p-mode frequency difference between 2000.4
and 1997.4.
The lines correspond to various distributions of the averaged magnetic
field, shown in the middle panel.   The solid and
dotted lines are within the error bars in the top
panel. The bottom panel shows the relative temperature decreases
required to cause similar frequency shifts as the corresponding magnetic
field increases}
\end{figure}

In Fig. 4 we show two examples of the field's changing behavior
that would be consistent with the observed $\gamma$'s, and
compare them with two cases that are clearly inconsistent. One of
the two inconsistent cases is a depth independent increase, and the other is
an example of the field gradually increasing to about 3 kG at 8Mm.
In all four examples, we used $\beta=-1$. We see that indeed
the $\gamma_p(\nu)$ provides a strong constraint on the
localization of the source of frequency changes, but clearly not a
unique answer.
For the two fitted cases, the inferred values of $B_b$
are 290 and 250 G. Corresponding values of
$B_{\rm ph}\equiv\Delta<\!B\!>_{\rm rms}(0)$ are
62 and 94 G.
An equally good fit was obtained with the choice
$\beta=1/3$. Data on the three models of the magnetic field
change fitting $\Delta\gamma_p(\nu)$ data are given in Table 2.
The result for $\beta=1/3$ is not significantly different from
that found by Goldreich et al. (1991). To explain the p-mode frequency
increase between minimum and maximum, we require an increase of the
rms magnetic field  growing from 0.2 kG in the photosphere to 0.84
kG at 4.25 Mm. The corresponding numbers of Goldreich et al. are
0.25 and 1 kG.

In the bottom panel of Fig. 4, we plot the relative temperature
changes, which gives the same local contributions to
$\Delta\gamma_p$ as the corresponding changes in the magnetic
field. We see that the required change of temperature is
unacceptably large in the atmospheric layers. However, in
sub-photospheric layers we cannot exclude the rms $\Delta T/T$ at
the $10^{-3}$ level. Such a temperature decrease would be a
significant contributor to the observed frequency increase.
Br\"uggen and Spruit (2000) argue that one expects a lower subsurface
temperature from an increasing magnetic field, and that
the effect should be searched for by means of helioseismology. A
contribution from temperature decrease would lower the requirement
for the magnetic field increase in the sub-photospheric layers.

Yet another potential contributor to the frequency increase is a
decrease in the turbulent velocity. Roughly, the relative change in
the turbulent velocity, $\Delta v_t/v_t=q$, has the same effect as a relative
temperature change $\Delta T/T=0.5q {\cal M}^2$ , where ${\cal M}$ is
the turbulent Mach number. In the
sub-photospheric layers, ${\cal M}$ is in the 0.1--1 range. Thus, the
effect may be significant, and we may expect a decrease in $v_t$,
with increasing activity,
because the magnetic field should inhibit convection.

\subsection{Shrinking or expanding of the outermost layers}

\begin{table}
\caption{Inference from p-mode frequency changes between 1997.4 and 2000.4}

\medskip

\begin{tabular}{ccccccc}
\hline
$\beta$ &  $\lambda$ & $D_b$ & $D_t$& $B_b[G]$& $B_{\rm ph}[G]$ &
$(dR/dt)_{\rm ph}$\\
\hline
 -1 & 0.575 & 3.00 & -0.485 & 29 & 62 &-1.3 km/y\\
 -1 & 1.15 & 1.27 & 0 & 25 & 94 & -1.1 km/y\\
 1/3 & 0.623 & 4.25 & -0.485 & 84 & 200 & 2.3 km/y\\
\hline
\end{tabular}
\end{table}
In Table 2, we provide the values of the contribution to the rate
of the photospheric radius change due to the magnetic field
increase inferred from the $\gamma_p$ changes. We emphasize
that the rate does not refer to photosphere but to the mass point
corresponding to the unperturbed (solar minimum) photosphere and
that the value does not include the part that was inferred from
f-mode frequency changes.

The solar photosphere is defined as a surface of specified optical
depth $\tau_{\rm ph}=M_{\rm ph}\bar\kappa$, where 
$M_{\rm ph}$ is column-mass depth and $\bar\kappa$ is
the mean opacity in the atmosphere, or, which is
closely related, the place where the local temperature
equals the effective temperature. Thus, if we want to assess the
rate of movement of the photosphere, we have to take into account a
possible change in $\bar\kappa$. To keep $\tau_{\rm ph}$
unchanged, an additional radius shift of
$-\Delta\bar\kappa(dr/d\bar\kappa)_{\rm ph}$ is needed. Hence, the
rate of the photosphere's change may be assessed as
\begin{equation}
{d R_{\rm ph}\over dt}=\left({d R\over dt}\right)_{\rm ph}-
\left({d\bar\kappa\over dt}{dr\over d\bar\kappa}\right)_{\rm ph},
\end{equation}
An estimate shows that the second term may not be negligible if
$(\Delta T/ T)_{\rm ph}\sim10^{-3}.$

The connection between $R_{\rm ph}$ and the solar disk 
radius, $R_d$, determined from the inflection point in the
limb-darkening function was discussed recently by Brown and
Christensen-Dalsgaard (1998). They find $R_d-R_{\rm
ph}\approx500\mbox{ km}$. Again a $10^{-3}$ temperature
perturbation within the atmosphere may be significant at the level
of the radius changes discussed here. Thus, the difference between the
solar radius variations inferred by means of seismology and
photometry has to be kept in mind when a detailed comparison is
made.

The total value of $(d R/ dt)_{\rm ph}$ may be estimated as the
sum of -1.5 km/s inferred from the f-mode data and one of the
values inferred from p-modes data shown in Table 2. These are
model dependent. We do not expect that by including the nonmagnetic
contributors to $\Delta\gamma_p$,
we would infer rates significantly beyond the range of values
quoted in this table. We note that the net effect may imply both
contraction and expansion. Possible net values of $(d R/ dt)_{\rm ph}$ 
range
from -3 to 1 km/y.

Finally, we  point out that there is no contradiction
between our inferences from f- and p-mode frequency changes. The
effect of the field increases needed to account for the
$d R_f/dt$ value have a negligible effect on p-modes frequencies, if
the outward decrease of $d\!<B\!>_{\rm rms}/dt$ from the bottom of
the convective zone is steep enough.

\section{Conclusions}

Results of our analysis of f-mode frequency confirm the evidence,
first found by Antia and Basu (2000), for a contraction of the
sun's outer layers during the rising phase of the magnetic
activity. The rate we determine is 1.5 km/y and is only somewhat
different than
found by our predecessors. We pointed out, however, that there may
be another interpretation for the observed frequency variation.
Further, we stressed that the rate does not refer to the surface
radius, but to the layer at 4 -- 8 Mm depth below the photosphere.
In spite of the fact that the dispersion relation for high-degree f-modes
approaches that for the surface gravity waves, the two types of modes
are essentially different. While the latter are discontinuity modes
which see the same gravity for each horizontal wave number, the f-modes
see different effective gravities depending on $\ell$.

The rate of shrinking is most easily explained as resulting from the
rise of the radial component of the random magnetic field beneath a depth
of 8 Mm. There is an integral constraint on the magnetic field which
may be the most important finding from the data on f-mode
frequency changes. To account for the shrinking rate, we need an
increase in the radial component of the random magnetic field with a
modest annual rate. An isotropic increase would imply an
expansion in the f-mode region.  We pointed out that this new constraint 
is likely to be in conflict with the much larger change of the interior
field inferred  by Dziembowski et al.( 2000) from the inversion of
p-mode frequency changes.

The p-mode frequency change may be accounted for in terms of
magnetic field changes. Our analysis was based on the formalism
of Goldreich et al. (1991), and we
found similar implications regarding the required field increase as
these authors, who analyzed BBSO data from the previous solar maximum.
In particular, the increase must be larger below the photosphere
than in the atmosphere, if this is the sole effect causing
p-mode frequency changes. We pointed out, however, that a
temperature decrease and/or decrease of turbulent velocity in
sub-photospheric layers could be significant contributors to the frequency
decrease. Depending on the field anisotropy,
the changes in the outermost layers may
lead to additional shrinking or to net expansion.

Our estimated
rates of radius change during the rise of cycle 23
range from -3 to 1 km/y.
This differs from the rate of about 5.9$\pm$0.7 km/y determined by
Emilio et al. (2000) from the direct radius measurements
based on SOHO/MDI intensity data. Perhaps
the difference may be explained by the difference between $dR_d/dt$ and
our $(dR/dt)_{\rm ph}$.
Both results, however, imply
a negligible contribution of the radius change
to the solar irradiance variations. Furthermore, the two estimates
of the radius change between maximum and minimum activity are
by two orders of magnitude less than found by N\"oel (1997) from
his measurements with the astrolabe of Santiago. He finds the difference
between the 1991 (previous maximum) and 1996 radii
which is exceeding 700 km.
The data from the Solar Diameter Monitor
(Brown \& Christensen-Dalsgaard, 1998) are inconsistent with such
large variations, although there is a hint of possible radius increase
during 1987 of some 30--40 km.
On the other hand, a theoretical constraint on radius
given by Spruit (1994) is even tighter than than that
from helioseismology. The number he quotes for the maximum to
minimum difference is $2\times10^{-7}R_\odot=0.14$ km.

We thank an anonymous referee for careful and detailed comments that
improved our paper.

 SOHO is a project of international cooperation between ESA and
NASA. This research was supported by the SOI-MDI NASA contract
NAG5-8878 at Stanford University, and partially supported by
NASA-NAG5-9730, NASA-NAG5-9543 and  NSF-ATM-97-14796 grants to Big
Bear and KBN-2-P03D-014-14 to W.A.D..
\newcommand{\Y}{Y_\ell^m}
\newcommand{\bn}{\mbox{\boldmath{$\nabla$}}}
\newcommand{\bxi}{\mbox{\boldmath{$\xi$}}}
\newcommand{\F}{\mbox{\boldmath{${\cal F}$}}}
\begin{center}
APPENDIX
\end{center}
We assume the Cowling approximation and write the equation for adiabatic
oscillation in the following form
$$ \rho\omega^2\bxi=\bn P'+\rho'g{\bf e}_r\equiv{\bf F}\bxi.\eqno(A1)$$
The notation here is a standard one and does not require explanation.
The variational expression for eigenfrequencies is
$$\omega^2={\int d^3{\bf x}\bxi^*\cdot{\bf F}\bxi\over\int d^3{\bf x}
\rho\vert\bxi\vert^2}\equiv{K\over I}.\eqno(A2)$$
The f-modes are nearly incompressible. Thus, for the approximate $\bxi$ to be
used in this expression, we assume
$$\bn\cdot\bxi=0.\eqno(A3).$$
Then for the Eulerian perturbation of density and pressure, we have
$$P'=g\rho\xi_r\quad\mbox{ and }\quad\rho'=-{d\rho\over dr}\xi_r.\eqno(A4)$$
We express in a standard way the displacement eigenvector in terms of
the spherical harmonics,
$$\bxi=[y(r){\bf e}_r+z(r)\bn]\Y.\eqno(A5)$$
With this expression (A3) becomes
$${dy\over dr}+2{y\over r}-L^2{z\over r}=0\eqno(A5)$$
and the integrals in (A2) become
$$I=\int_0^R(y^2+L^2z^2)\rho r^2dr\eqno(A6)$$
and
$$K=\int_0^R\left[2L^2yz+\left({d\ln g\over d\ln r}-2\right)y^2\right]
{g\over r}\rho
r^2dr,\eqno(A7)$$
where we made use of (A4).
We may use $${d\ln g\over d\ln r}=-2,$$
because for modes considered here the logarithmic derivative of the local
mass, $M_r$, is less than $10^{-2}$ in the layers contributing to $I$ and $K$,
which implies less than a $10^{-4}$ fractional contribution to frequencies.
From the ratio of radial to horizontal component of
(A1), we obtain approximately $${y\over z}=L^2{z\over y}$$
and, taking into account the inner boundary condition, $y=Lz$.
Now, we have from (A6)
$$ I= 2\int_0^R y^2\rho r^2dr\eqno(A8)$$
and from (A7)
$$K=2(L-2)\int_0^R y^2{g\over r}\rho r^2dr\eqno(A9).$$
Eqs.(2) and (3) follow immediately from (A2), (A8), and (A9) upon
setting $M_r=M$ and $\omega=2\pi\nu$.

\end{document}